\documentclass{osa-article}  
\journal{boe}
\articletype{Research Article}
\usepackage[utf8]{inputenc} 
\usepackage[T1]{fontenc}
\usepackage{amsmath,amsfonts}
\usepackage{graphicx, booktabs, multirow}
\usepackage{siunitx}
\usepackage{units}
\usepackage[colorlinks=true]{hyperref}
\usepackage{nicefrac}
\usepackage{float, enumerate, adjustbox}
\usepackage{stackengine}
\usepackage{tabularx}
\usepackage{subfig}
\usepackage{graphicx}
\usepackage{algorithm}
\usepackage[noend]{algpseudocode}
\usepackage{xcolor}
\usepackage{nomencl}
\usepackage{soul}
\DeclareMathOperator*{\argmin}{arg\,min}
\providecommand{\keywords}[1]
{
  \small	
  \textbf{\textit{Keywords--}} #1
}

\makenomenclature

\begin{document} 
\pagestyle{plain}

\title{Band selection for oxygenation estimation with multispectral/hyperspectral imaging}

\author{Leonardo Ayala,\authormark{1,7,*} Fabian Isensee,\authormark{2,6,*} Sebastian J. Wirkert,\authormark{1} Anant S. Vemuri,\authormark{1}
 Klaus H. Maier-Hein,\authormark{2,7,8} Baowei Fei,\authormark{3,4,5} Lena Maier-Hein,\authormark{1,7,8}}

\address{\authormark{1}Division of Computer Assisted Medical Interventions (\textsc{CAMI}), German Cancer Research Center (\textsc{DKFZ}), Heidelberg, Germany\\
\authormark{2}Division of Medical Image Computing (\textsc{MIC}), German Cancer Research Center (\textsc{DKFZ}), Heidelberg, Germany\\
\authormark{3}Department of Bioengineering, The University of Texas at Dallas, Richardson, Texas, United States\\
\authormark{4}Advanced Imaging Research Center, The University of Texas Southwestern Medical Center, Dallas, Texas, United States\\
\authormark{5}Department of Radiology, The University of Texas Southwestern Medical Center, Dallas, Texas, United States\\
\authormark{6}Applied Computer Vision Lab, Helmholtz Imaging\\
\authormark{7}Medical Faculty, Heidelberg University, Heidelberg, Germany\\
\authormark{8}Faculty of Mathematics and Computer Science, Heidelberg University, Heidelberg, Germany\\
\authormark{*}Authors contributed equally\\}
\email{l.menjivar@dkfz-heidelberg.de}
\email{l.maier-hein@dkfz-heidelberg.de}

\begin{abstract}
Multispectral imaging provides valuable information on tissue composition such as hemoglobin oxygen saturation. However, the real-time application of this technique in interventional medicine can be challenging due to the long acquisition times needed for large amounts of hyperspectral data with hundreds of bands. While this challenge can partially be addressed by choosing a discriminative subset of bands, the band selection methods proposed to date are mainly restricted by the availability of often hard to obtain reference measurements. We address this bottleneck with a new approach to band selection that leverages highly accurate Monte Carlo (MC) simulations. We hypothesize that a so chosen small subset of bands can reproduce or even improve upon the results of a quasi continuous spectral measurement. We further investigate whether novel domain adaptation techniques can address the inevitable domain shift stemming from the use of simulations. Initial results based on \textit{in silico} and \textit{in vivo} experiments suggest that 10-20 bands are sufficient to closely reproduce results from spectral measurements with 101 bands in the 500-700 nm range. The investigated domain adaptation technique, which only requires unlabeled \textit{in vivo} measurements, yielded better results than the pure \textit{in silico} band selection method. Overall, our method could guide development of fast multispectral imaging systems suited for interventional use without relying on complex hardware setups or manually labeled data.
\end{abstract}



\mbox{}
\nomenclature[01]{$r(\lambda)$}{Simulated tissue reflectance}
\nomenclature[02]{$r_k$}{Simulated reflectance adapted to camera system band $k$}
\nomenclature[03]{$(I_k)_{(i,j)}$}{Reflectance measurement at image pixel $(i,j)$ and camera band $k$}
\nomenclature[04]{$X_g$}{Generic tissue reflectance simulations}
\nomenclature[05]{$X_c$}{Colon tissue reflectance simulation}
\nomenclature[06]{$X_{ac}$}{Tissue reflectance simulations selected from generic tissue $X_g$ to resemble colon tissue reflectance $X_c$}
\nomenclature[07]{$X_{am}$}{Tissue reflectance simulations selected from generic tissue $X_g$ to resemble \textit{in vivo} mouse tissue}
\nomenclature[08]{$\mathbf{r_{nn}}$}{Nearest neighbour simulated reflectance to measured reflectance}

\printnomenclature

\section{Introduction}
\label{sec:intro}
Multispectral and hyperspectral imaging (MSI/HSI) could be useful for many applications in surgery, including tumor detection and perfusion monitoring \cite{lu_medical_2014,clancy2020,ayala2019,dietrich2021}. Acquisition of many spectral bands, however, leads to long imaging times and/or low resolution, hampering widespread adoption of the technique. To overcome this issue, current research focuses on reducing the number of recorded bands. Yet, the methods proposed are not capable of considering both the target domain (e.g. liver surgery) and the specific task (e.g. oxygenation or blood volume fraction monitoring) when selecting bands \cite{wood_optimal_2008, du_band_2007}.

In this paper, we propose a method that can provide task-specific and target domain-specific band selection. The generic applicability of our approach is achieved with highly generic Monte Carlo (MC) tissue simulations that aim to capture a large range of optical tissue parameters potentially observed during surgical interventions. The adaptation of the model to a specific clinical application is based on label-free \textit{in vivo} hyperspectral recordings using a published approach to multispectral domain adaptation \cite{huang_correcting_2007}. The bands are selected based on their performance to estimate task-specific physiological parameters.

In Section \ref{sec:soa} we give a detailed overview of the current state-of-the-art. In Section \ref{sec:method} we describe the method outlined above in more detail. Furthermore, we present in Section \ref{sec:exp} the validation of the method \textit{in silico} and with \textit{in vivo} data from \cite{lu_framework_2015}. We finalize the manuscript by including a discussion of the findings in Section \ref{sec:dscussion}.

\section{State-of-the-art in multispectral band selection}
\label{sec:soa}

A large body of work in band selection is available in the fields of remote sensing  \cite{sheffield_selecting_1985, beauchemin_statistical_2001, guo_band_2006, chang_constrained_2006, sarhrouni_dimensionality_2012}, food safety \cite{du_band_2007} and histopathology \cite{paul_regenerative_2015}. This state-of-the-art overview will be restricted to algorithms related to interventional imaging, as summarized in Table \ref{tab:soa_bs}.

Band selection is closely linked to variable/feature selection, in which algorithms can be grouped roughly in \textit{filter}, \textit{wrapper} and \textit{embedded} methods \cite{guyon_introduction_2003}. \textit{Filter} methods determine the best features as an independent preprocessing step, which is unaware of the target task (e.g. oxygenation estimation). \textit{Wrapper} methods use the downstream processing pipeline to search for feature combinations that maximize the target performance metric for a given task (e.g. accuracy of tumor detection). Finally, \textit{embedded} methods are incorporated into the training process of a machine learning algorithm, where feature selection becomes part of the model construction. While \textit{filter} methods can rely on unsupervised correlation analysis, \textit{wrapper} and \textit{embedded} methods usually need labeled training data. Note that band selection is different from dimensionality reduction methods such as principal component analysis (PCA), which reduce the dimensionality by finding a subset composed of combinations of bands/features. This does not reduce imaging time, since all bands first have to be recorded to compute the subset found by PCA.

\begin{table}[h!]
\centering
\caption{Overview of relevant band selection methods. \textit{Selected bands} represent the optimal number of bands reported by the authors. Our proposed method suggests a subset of 10 out of 101 bands for the \textit{in-silico} oxygenation experiments, but  can be applied to a wide range of tasks.}
\label{tab:soa_bs}
\begin{tabular}{p{2cm} p{1.5cm}p{1.3cm}p{1.2cm}p{1.3cm}p{3cm}p{1cm}}
\hline
Author       & Modality & label-free & Method   & Data      & Application    & selected bands            \\ \hline \hline
Asfour \cite{Asfour2018}        & HSI      & yes        & \textit{wrapper}   & \textit{ex vivo}   & tissue visualization    & 4/151  \\
Gu \cite{gu_image_2016}        & MSI      & yes        & \textit{filter}   & \textit{in vivo}   & tissue visualization    & 6/16  \\
Han \cite{han_vivo_2016}      & HSI      & no         & \textit{filter}   & \textit{in vivo}   & cancer localization    & 5/28      \\
Marois \cite{marois_optimal_2018} & MSI & yes & \textit{wrapper} & \textit{in silico} & functional imaging      & 6\\
Nouri \cite{nouri_efficient_2014, nouri_hyperspectral_2016}     & HSI      & yes        & \textit{filter}   & \textit{in vivo}   & tissue visualization    & 3/141 \\
Preece \cite{preece_spectral_2004}   & MSI      & yes        & \textit{wrapper}  & \textit{in silico} & functional imaging    & 3     \\
\rowcolor{lightgray}
Proposed & HSI      & yes        & \textit{wrapper}   & \textit{in vivo}   & functional imaging      & 10/101 \\ 
Wirkert \cite{wirkert_endoscopic_2014} & MSI      & yes        & \textit{filter}   & \textit{in vivo}   & functional imaging    & 8/20   \\
Wood \cite{wood_optimal_2008}      & MSI  & no         & \textit{wrapper}  & \textit{ex vivo}   & cancer localization    & 3/16  \\

\hline
\end{tabular}
\end{table}

State-of-the-art methods shown in Table \ref{tab:soa_bs} can be classified into three categories depending on the task being addressed: a) Cancer localization, b) tissue visualization and c) functional imaging.

\subsection*{a) Cancer localization}
Contributions in the area of cancer localization focus on selecting bands that maximize the performance of a classifier. Some of the most prominent contributions include the work of Wood et al. \cite{wood_optimal_2008}, who selected the bands which maximize the Area Under the Curve (AUC) of a Na\"ive Bayesian classifier's Receiver Operating Characteristic (ROC). More specifically, they used a greedy algorithm to remove the bands which least contribute to the AUC while featuring a low AUC if evaluated alone. The algorithm was evaluated on phantoms and stained lung cancer biopsies, and the authors concluded that the use of three wavelengths provides essentially as much information as the use of all sixteen.

Another contribution in this area focuses on gastric tumor localization. Han et al. \cite{han_vivo_2016} selected bands which maximized the symmetric class-conditional Kullback-Leibler (KL) divergence between disease and normal tissue reflectance spectra. A suboptimal, greedy search algorithm \cite{du_band_2007} was used for this purpose because the possible number of band subsets grows exponentially with the number of bands. This algorithm iteratively adds bands to a selection set based on the largest incremental increase of the KL measure. The algorithm was evaluated on 12 patients and a total of 21 colorectal tumors, and the authors concluded that 5 out of 28 bands are useful for outlining tissue disease regions.

\subsection*{b) Tissue visualization}
Several approaches address the problem of selecting the bands that improve the contrast between normal and abnormal tissue. Among these, Gu et al. \cite{gu_image_2016} selected three bands to discriminate gastric abnormalities from benign tissue. They aimed to replace RGB images with the selected bands. Their algorithm first selects the band with the highest variance in the visible range from a set of 27 bands. The algorithm then adds subsequent bands iteratively with the criterion of minimizing mutual information between the current set of bands and the previous selection. The set of bands was optimized using 29 images from 12 patients with gastric abnormalities. The authors concluded that 3 selected bands increase contrast compared to RGB images.\\
In a similar approach, Nouri et al.  \cite{nouri_efficient_2014, nouri_hyperspectral_2016} inspected a number of unsupervised, thus label-free, band selection algorithms. The algorithms, originating from the remote sensing community, were evaluated within the context of hyperspectral ureter surgery. Instead of identifying bands which separate a certain class, the authors aimed to identify three bands for better visualization (instead of RGB) to present to the surgeon. They evaluated the competing algorithms by several contrast and entropy metrics and assessed how different sets of bands can differentiate structures such as the ureter and adipose tissue. They found that the three best wavelengths to discriminate ureter tissue are situated in the near infrared region and that the Sheffield Index preserves a maximum of information.\\
In another work, Saiko et al. \cite{saiko2020} studied the optimal number of bands to increase the contrast of tissue structures. They linked the image contrast ratio with optical tissue parameters based on the two-flux Kubelka-Munk model. They reported that bands around 550nm maximize the contrast.
Furthermore, Asfour et al. \cite{Asfour2018} investigated the optimal bands to improve the visualization of atrial ablation lesions. They evaluated the performance of their algorithm in subsets of 2, 3 and 4 bands based on supersets containing 151 and 31 equidistant bands in the wavelength range between 420nm and 720nm. The authors concluded that 4 bands can be used to enhance the contrast of ablated atrial tissue.

\subsection*{c) Functional imaging}
Functional imaging methods in the medical field are techniques used to assess the state of metabolism, blood flow, chemical composition, etc. either spatially or temporally within organs. In this context, band selection methods aim to select a subset of bands while maintaining good performance in at least one of such tasks. The method proposed by Wirkert et al. \cite{wirkert_endoscopic_2014} selects bands in a completely unsupervised manner by maximizing the differential entropy contained in the selected subset of bands. More specifically, the algorithm selects the subset of bands with the highest determinant of the bands' covariance matrix. Assuming an underlying normal distribution, this determinant is proportional to the differential entropy, and thus also to the information contained in the subset of bands. The authors tested the method in an \textit{in vivo} porcine setting and concluded that a selection of 8 bands leads to blood oxygenation values close to the baseline generated with 20 bands.

In contrast to this approach, Marois et al. \cite{marois_optimal_2018} selected bands by analyzing the absorption matrix, assuming oxygenation can be computed from the modified Beer-Lambert's law and leveraging a wavelength-dependent path length factor. They chose bands that maximize the product of the singular values of the absorption matrix, arguing that this maximizes the orthogonality of the fitted spectra. They then evaluated the quality of the chosen bands in an \textit{in silico} setting in the visible and infrared regions and reported the root mean squared error of the estimated concentrations of different chromophores. Ultimately, they reported six optimal wavelengths for estimating the concentration of water, lipids, oxygenated and deoxygenated hemoglobin.

Preece et al. \cite{preece_spectral_2004} selected bands with a different idea. First, Kubelka-Munk light transport theory-based simulations were created for assessing pigmentation of human skin. After adapting these simulations to a set of virtual filters, a genetic algorithm was employed to find the best subset for estimating papillary dermis thickness and blood/melanin content. The authors ensured that the bands could invert the parameters uniquely by differential-geometric reasoning. 
Incorporation of ground truth from simulations allowed them to circumvent problems related to references from real data as mentioned in \cite{lu_framework_2015}. The authors concluded that three bands selected according to their method lead to better results than RGB, but that RGB leads to reasonable results in the investigated context.\\

\subsubsection*{Novelty of our contribution}
Our method differs from previous work in several key aspects.
The methods proposed by Wood, Han and Nouri \cite{wood_optimal_2008, han_vivo_2016, nouri_efficient_2014} require labeled data, e.g. according to a malignancy classification; in contrast, we leverage unlabeled \textit{in vivo} data. The \textit{filter} methods developed by Nouri, Gu and Wirkert{\cite{nouri_hyperspectral_2016, gu_image_2016, wirkert_robust_2016}} optimize band selection based on the maximization of a non-specific measure such as amount of information, while we developed a method that can be adapted to  specific tasks. The approach by Preece et. al. \cite{preece_spectral_2004} requires specific knowledge about the tissue composition and can not be adapted to real tissue images, which we overcome by employing domain adaptation techniques. Furthermore, while the aforementioned work relied entirely on simulations, we also evaluate band selection in an \textit{in vivo} context, ensuring that both simulations and band selection results are closer to a real surgical scenario.

\section{Method}
\label{sec:method}
\begin{figure}[h!bt]
  \begin{center}
  \includegraphics[width=0.8\textwidth]{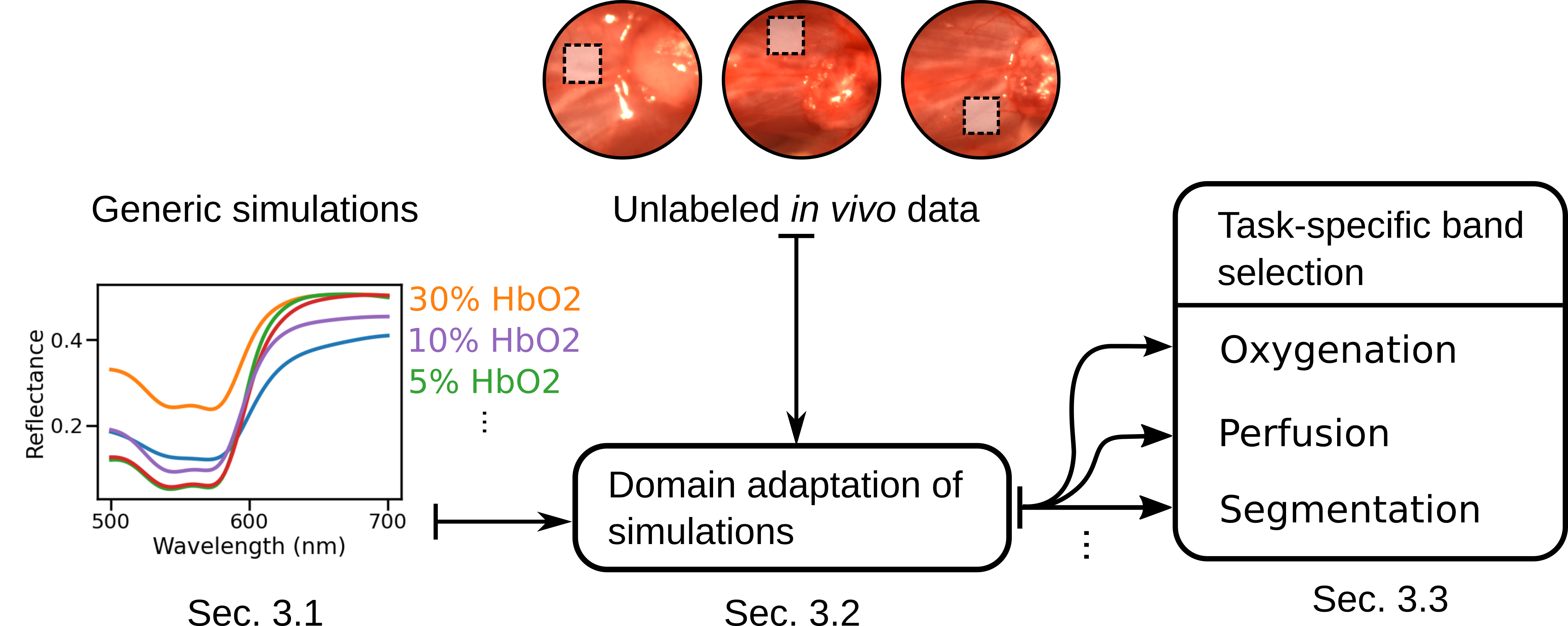}
  \end{center}
  \caption{Overview of our approach. Generic simulations are adapted using unlabeled hyperspectral measurements from the target domain. The resulting domain-specific simulations are the basis for the subsequent task-specific band selection. \label{fig:concept}}
\end{figure}

The proposed method chooses bands that optimize a performance criterion with respect to a specific task (e.g. oxygenation monitoring) and domain (e.g. colonoscopy) without the need for annotated reference measurements. At the core of the method is a simulated generic data set with ground truth knowledge of optical and functional parameters (e.g. oxygenation, Sec. \ref{sec:reference}). This dataset is leveraged by the proposed domain adaptation technique (Sec. \ref{sec:da}) and thus serves as foundation for the actual band selection algorithm (Sec. \ref{sec:bandselection}). Fig. \ref{fig:concept} summarizes the proposed method.

\subsection{Reference data generation}
\label{sec:reference}
Generic simulated data is generated as described in \cite{wirkert_physiological_2017}, briefly revisited here. Tissue is assumed to be composed of three infinitely wide slabs. Each of these slabs is defined by values of blood volume fraction $v_{\textrm{hb}}$, blood oxygen saturation $s$, reduced scattering coefficient at 500nm $a_{\text{mie}}$, scattering power $b_\text{mie}$, anisotropy $g$, refractive index $n$ and layer thickness $d$. Values from literature \cite{jacques_optical_2013} are used to create MC simulations; these values include: Extinction coefficients of hemoglobin $\epsilon_{\text{Hb}}$ and deoxyhemoglobin $\epsilon_{\text{HbO2}}$, absorption $\mu_a$ and scattering $\mu_s$ coefficients. A Graphics Processing Unit (GPU) accelerated version \cite{alerstam_next-generation_2010} of the popular Monte Carlo Multi-Layered (MCML) simulation framework \cite{wang_monte_1992} was chosen to generate spectral reflectances. The ranges from which the parameters are uniformly sampled as well as general simulation parameters are summarized in Table \ref{tab:generic}. 

\begin{table}[h!]
\centering
\caption{The simulated ranges of physiological parameters, and their usage in the simulation setup. Here $v_{\textrm{Hb}} \lbrack \% \rbrack$ represents the blood volume fraction, $s$ the blood oxygen saturation, $a_{mie}$ the reduced scattering coefficient at $500 nm$, $b_{mie}$ the scattering power, $g$ the tissue anisotropy, $n$ the refractive index and $d$ the tissue thickness. All parameters have been uniformly sampled within the specified range.\label{tab:generic}}

\begin{tabular}{p{1.1cm} c c c c c c c}
\hline
      & $v_{\textrm{Hb}} \lbrack \% \rbrack$           & $s \lbrack \% \rbrack$  & $a_{\text{mie}} \lbrack \si{cm^{-1}} \rbrack$ & $b_\text{mie}\unit{[a.u.]}$         & $g\unit{[a.u.]}$      & $n\unit{[a.u.]}$         & $d \lbrack\unit{cm}\rbrack$      \\ 
\hline
layer 1: & $0-30$ & $0-100$ & $5-50$ & $0.3-3$ & $0.80-0.95$ & $1.33-1.54$ & $0.002-0.2$  \\ 
layer 2: & $0-30$ & $0-100$ & $5-50$ & $0.3-3$ & $0.80-0.95$ & $1.33-1.54$ & $0.002-0.2$\\
layer 3: & $0-30$ & $0-100$ & $5-50$ & $0.3-3$ & $0.80-0.95$ & $1.33-1.54$ & $0.002-0.2$\\
\hline \hline
\multicolumn{8}{l}{$\mu_a(v_{\text{Hb}}, s, \lambda) = v_{\text{hb}} (s\cdot \epsilon_{\text{HbO2}}(\lambda) + (1-s)\cdot\epsilon_{\text{Hb}}(\lambda))\cdot \ln (10)\cdot150\si{g.L^{-1}}\cdot(\num{6.45e4}\si{g.mol^{-1}})^{-1}$} \\
\multicolumn{8}{l}{$\mu_s(a_{\text{mie}}, b, \lambda) = \frac{a_{\text{mie}}}{1-g}(\frac{\lambda}{500\si{nm}})^{-b_{\text{mie}}}$}  \\ \hline                 
\multicolumn{8}{l}{simulation framework: GPU-MCML\cite{alerstam_next-generation_2010}, $10^6$ photons per simulation}                                                \\                  
\multicolumn{8}{l}{simulated samples: \num{5.5e5}}                                                \\
\multicolumn{8}{l}{sample wavelength range $[\lambda_{min}-\lambda_{max}]$: 300\si{nm}-1000\si{nm}, stepsize 2\si{nm}}\\
\hline
\end{tabular}
\end{table}

The wavelength range $[\lambda_{min}-\lambda_{max}]$ is large enough for adapting the simulations to cameras operating in the visible and near infrared optical range. The simulated spectral reflectances $r(\lambda)$ are transformed to the reflectance camera measurement $r_k$ at band $k$ according to Eq. \ref{eq:camera_reflectance}. 
\begin{equation}
r_k=\frac{\int_{\lambda_{\text{min}}}^{\lambda_{\text{max}}}o(\lambda)\cdot l(\lambda)\cdot f_k(\lambda)\cdot r(\lambda)\,\text{d} \lambda }
{\int_{\lambda_{\text{min}}}^{\lambda_{\text{max}}}o(\lambda)\cdot l(\lambda)\cdot f_k(\lambda)\,\text{d} \lambda } \in [0,1]
\label{eq:camera_reflectance}
\end{equation}
Here, $f_k(\lambda)$ characterizes the $k$th  optical filter response of the camera, $l(\lambda)$ represents the relative irradiance of the light source and $o(\lambda)$ describes other parameters of the optical system such as camera quantum efficiency and transmission of additional optical elements (optical lenses, light guides, etc.). Multiplicative Gaussian noise is added to the simulated reflectances to account for camera noise and inaccuracies arising from modelling tissue as homogeneous layered structures. 
See Sec. \ref{sec:data_preparation} for more details on the specific camera parameters used for the data analysis.

\subsection{Adapting to the target domain}
\label{sec:da}
The model from Sec. \ref{sec:reference} describes a generic tissue, which encompasses virtually all tissue types that have hemoglobin as main light absorber. However, for a given domain such as cancer localization, many of the generated simulations might be irrelevant and thus bands selected based on these simulations might be suboptimal. Domain adaptation techniques can ensure that the simulations match the target domain more closely \cite{pan_survey_2010}. In this manuscript, kernel mean matching (KMM)\cite{huang_correcting_2007}, a state-of-the-art domain adaptation technique, is used to automatically assign a weight to each simulation according to how closely they mimic real measurements taken from the target domain. More specifically, the following equation is minimized:
\begin{equation}
\label{eq:kmm}
\arg \min_{\boldsymbol{\beta}} \left\Vert \frac{1}{n} \sum_{i=1}^{n} \beta_i \phi(\mathbf{r}_i) - \frac{1}{n'} \sum_{i=1}^{n'} \phi(\mathbf{I}_i) \right\Vert^2\quad\textrm{s.t.}\quad\beta_i \in [0, B] \quad\textrm{ and }\quad \left|\sum_{i=1}^n \beta_i - n\right| \le n\frac{B}{\sqrt{n}}
\end{equation}
Here, $\mathbf{r}_i$ represents one simulated reflectance spectrum belonging to a set of $n$ simulations and $\mathbf{I}_i$ represents one real measurement belonging to a set of size $n'$.
This objective function matches the empirical means of simulations $\mathbf{r}_i$ and real data $\mathbf{I}_i$ in a reproducing Kernel Hilbert Space induced by the kernel $K$. We set this function to the Gaussian radial basis function kernel: $\phi(\mathbf{r}_i, \mathbf{r_j}) = \mathrm{e}^{-\gamma (\left\|\mathbf{r_i} - \mathbf{r_j}\right\|_2)^2}$. In simple terms, the target of KMM is to weight Gaussians associated with each simulation to reproduce the distribution of the measurements as closely as possible. 
The first boundary condition ($\beta_i \in [0, B]$) limits the maximum influence of individual training samples, the second condition ensures that the term $\frac{1}{n} \sum_{i=1}^{n} \beta_i \phi(\mathbf{r}_i)$ is close to a probability distribution \cite{gretton_covariate_2009}. Unfortunately, equation \ref{eq:kmm} cannot be minimized directly due to the possible infinite dimension of $\phi$. The random kitchen sinks method \cite{rahimi_random_2007} finds an approximate representation  of $\phi(\mathbf{r}_i) \approx z(\mathbf{r}_i)$ by sampling from the Fourier transformation of a shift invariant kernel.
This enables solving the convex KMM objective function in its non-kernelized form using a standard optimizer.
Once the optimization has finished, the weights $\beta_i$ are used to sample with replacement from the original simulated data set to establish a data set which more closely resembles real tissue.

\subsection{Task-specific band selection}
\label{sec:bandselection}
Band selection aims at reducing the number of bands while maintaining high performance on a desired task.
In this manuscript, the chosen task, unless otherwise specified, is oxygenation estimation, and the metric employed is mean absolute error (MAE). The random forest-based oxygenation estimation method developed by Wirkert et. al. \cite{wirkert_robust_2016} is used to map from the reflectance spectrum to oxygenation values. This method uses MC simulations to learn this mapping and was found to be more accurate than the conventional linear Beer-Lambert approaches.
In principle, our method is compatible with any feature selection or regression and classification method. Note that for the sake of consistency with the feature selection literature, we use the term feature rather than band in this section, but they are equivalent in the context of this work. In this paper, several popular \textit{filter}- as well as \textit{wrapper}-based approaches are studied and compared.

In a supervised learning setting, \textit{filter} methods estimate the "usefulness" of features by computing a specific metric such as mutual information or conditional mutual information between features and between features and targets. In contrast, \textit{wrapper} methods use the performance of a machine learning model to select the best features. Since feature selection is an \textit{NP-hard} problem, greedy step-wise optimization algorithms are often used to restrict the search space that is explored with both \textit{filter} and \textit{wrapper} methods \cite{kohavi1997wrappers,brown2012conditional}.

\subsubsection*{Wrapper feature selection}
Here, sequential forward selection (SFS) and best first search (BFS) are used \cite{kohavi1997wrappers} as \textit{wrapper} feature selection methods. Both are greedy algorithms that construct feature subsets sequentially by optimizing a fitness criterion. SFS iteratively adds single features to a given subset by keeping the feature that most improves the fitness criterion. Features are added until the criterion has not improved in the last three additions. BFS not only adds but also removes features from an existing subset and widens the search space by keeping track of all explored subsets. If stuck in a local minimum, BFS falls back to previously explored high-scoring feature sets and continues the search from there. \textit{Wrapper} methods are flexible in the sense that one can use any fitness criterion to be optimized. In the experiments described in Sec. {\ref{sec:exp}} we employ our target metric, the MAE, as fitness criterion (Eq. {\ref{eq:wrappercriterion}}) and a random forest regressor as described in {\cite{wirkert_robust_2016}} for oxygenation estimation in all experiments.

\begin{equation}
    \label{eq:wrappercriterion}
    \argmin_{s_i\subset S} (l_{\textrm{MAE}}(y, f(x_{s_i})))
\end{equation}

Here, $S$ denotes the complete set of possible features, $s_i$ are the selected features (a subset of $S$), $l_{\textrm{MAE}}(y_i, f(x_{s_i}))$ is the MAE achieved with the subset $s_i$, features $x$ and regressor $f$ for the target variable $y$ (e.g. oxygenation). $l_{\textrm{MAE}}(y_i, f(x_{s_i}))$ is computed by running a threefold cross-validation on the training data while using only the selected subset of features {$s_i$}. The final subset of features (and thus feature set size) is selected by evaluating the obtained sets on a test dataset and selecting the one with minimum MAE. 

\subsubsection*{Filter feature selection}
Several mutual information-based selection criteria are employed for \textit{filter} feature selection: Minimum Redundancy Maximum Relevance (mRMR) \cite{peng2005feature}, Conditional Mutual Information Maximization (CMIM) \cite{fleuret2004fast}, Mutual Information based Feature Selection (MIFS) \cite{battiti1994using}, Interaction Capping (ICAP) \cite{jakulin2005machine}, Conditional Infomax Feature Extraction (CIFE) \cite{lin2006conditional} and Joint Mutual Information (JMI) \cite{yang2000data}. Brown et al. \cite{brown2012conditional} provide a thorough analysis of these criteria and identify a unifying theoretical framework from which they can be derived. Although these methods were originally developed for classification problems, they can be extended to the context of regression, provided that a mutual information estimator can be defined. Such an estimator must be able to compute the conditional mutual information between two feature subsets even when the target is not categorical.
For this purpose, Kraskov's nearest neighbor mutual information estimator \cite{kraskov2004estimating} is used to compute the mutual information between subsets. Furthermore, feature subsets are optimized by constructing them in a greedy step-wise manner as described in \cite{brown2012conditional}. \textit{Filter} feature selection algorithms return the order with which features are added to the subset but are incapable of returning the optimal number of features to be used. In this work, the optimal number of features was not selected explicitly. We instead report the result for all generated feature set sizes.

\subsection{Data normalization}
\label{subsec:data-normalization}
Distance and angle between the MSI camera and the tissue introduce multiplicative changes on the intensity of the measured spectrum. Band normalization is performed to remove dependency on these factors, which can typically not be controlled during an intervention. Recommendation of \cite{wirkert_robust_2016} are followed to normalize each reflectance spectrum by dividing it by its mean, followed by a negative logarithmic transformation ($-\log$) and a further $\ell_2$ normalization. Given a spectral image {$I \in \mathbb{R}^{N_x \times N_y \times N_s}$} of spatial dimensions $N_x \times N_y$ and number of channels {$N_s$}, {$I_k \in \mathbb{R}^{N_x\times N_y}$} represents one image channel {$k$} {$(k \in \{ 1,...,N_s \})$} and {$(I_k)_{(i, j)} \in \mathbb{R}$} represents the intensity value at the pixel position {$(i,j)$} ({$i \in \{ 1,...,N_x \}$}, {$j \in \{ 1,...,N_y \}$}). The normalized spectra $\bar{\mathbf{I}}_{(i,j)}$ can be computed as follows:

$$(I_k^{\log})_{(i,j)} = -\log\left(\frac{(I_k)_{(i,j)}}{(I_k)_{(i,j)}/(\sum_k (I_k)_{(i,j)})}\right)$$
$$\bar{\mathbf{I}}_{(i,j)} = \left\Vert\mathbf{I}^{\log}_{(i,j)}\right\Vert_2$$

The $-\log$ transformation and $\ell_2$ normalization are not strictly necessary from a theoretical standpoint, but empirically improved results. In addition, the data needs to be re-normalized whenever a different set of bands is used, because the normalization method works on a per-sample basis and uses all the available bands for normalization. As a side effect, this theoretically allows our approach to jointly optimize bands along with their normalization.

\section{Experiments and Results}
\label{sec:exp}
Experiments were performed to assess the band selection results both in an \textit{in silico}  and an \textit{in vivo}  setting. Sec. \ref{sec:data_preparation} describes the experimental setup with a specific focus on the parameters used to configure the algorithms. The purpose of the \textit{in silico} experiments was to assess the method in a quantitative manner (Sec. \ref{sec:insilico}). The \textit{in vivo} experiments assessed how well oxygenation estimated from many bands can be reproduced by bands selected with the proposed method (Sec. \ref{sec:invivo}).

\subsection{Experimental setup}
\label{sec:data_preparation}
To validate our method, generic MC reflectance simulations $X_g$ were created from the model described in table \ref{tab:generic}.  These generic simulations defined two disjoint sets: a training $X_g^{train}$ (450,000 simulations) and a test set $X_g^{test}$ (15,000 simulations). Bands were selected on a subset of 15,000 simulations selected randomly from the training set. The selections were evaluated by training on the entire training set and evaluating on the test set.

To mimic the camera used for the \textit{in vivo} application, Equation \ref{eq:camera_reflectance} was used to simulate multispectral bands every 2nm from 500nm to 700nm, each band represented by a Gaussian transmissions profile with a full width at half maximum (FWHM) of 20nm. Camera quantum efficiency and the optical system were assumed constant within the relatively narrow \textit{filter} bands. Gaussian multiplicative noise (5\%) emulated the camera noise. The range of 500nm to 700nm was chosen because the simulations and measurements (see sec.~\ref{sec:insilico}) did not match below 500nm and above 700nm. Furthermore, when comparing the simulated data, presented in sec.~\ref{sec:insilico}, to the measurements, a 14nm shift in the absorption spectrum was detected. To align simulations and measurements, the measurements were shifted by 14nm. The \textit{in vivo} measurements were transformed to reflectance by dark and flatfield correction. They were further blurred with a Gaussian kernel with a sigma of one pixel in the spatial domain to denoise the measurements.

Following the suggestions of Wirkert et al.\cite{wirkert_robust_2016}, the random forest was set to use 10 trees, a maximum depth of nine and a minimum number of samples per leaf of 10. As in \cite{wirkert_physiological_2017}, the KMMs parameter B (see eq. \ref{eq:kmm}) was set to 10, and the kernel function's parameter $\gamma$ was set to the median pairwise Euclidean sample distance determined on the training data.

\subsection{\textit{In silico} validation}
\label{sec:insilico}
Purpose of the \textit{in silico} experiments was to assess the different band selection techniques and the proposed domain adaptation approach in a quantitative manner.

\paragraph{Influence of band selection methods}

\begin{figure}[h!]
\centering
\begin{tabular}{cc}

\subfloat[Comparison of different \textit{filter} and \textit{wrapper} methods\label{fig:filterwrapper}]{%
  \includegraphics[width = 0.45 \columnwidth]{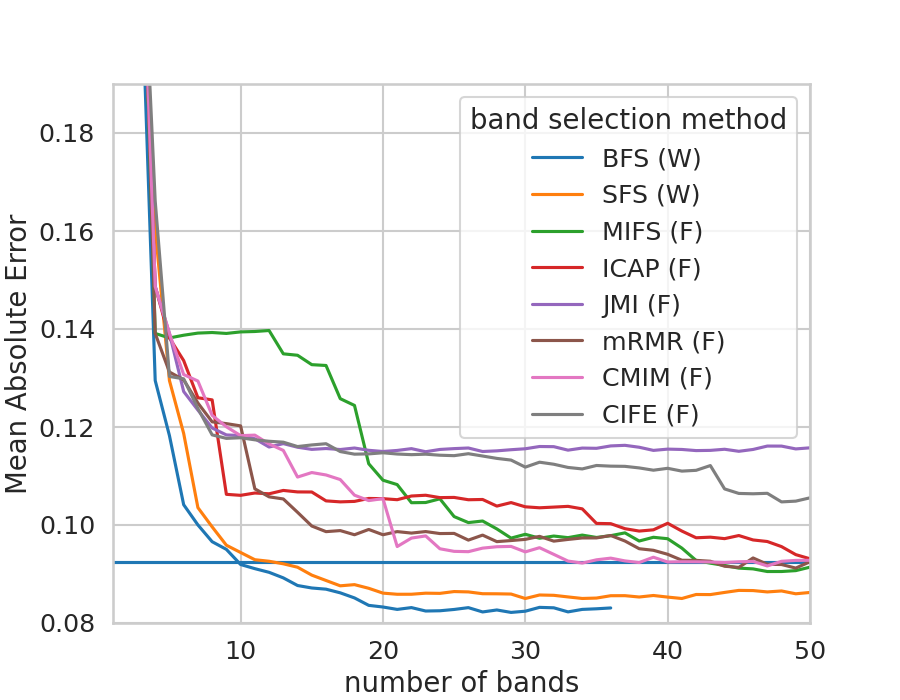}
}  &
\subfloat[Band selection results for selected methods\label{fig:bsmethods}]{%
  \includegraphics[width = 0.5 \columnwidth]{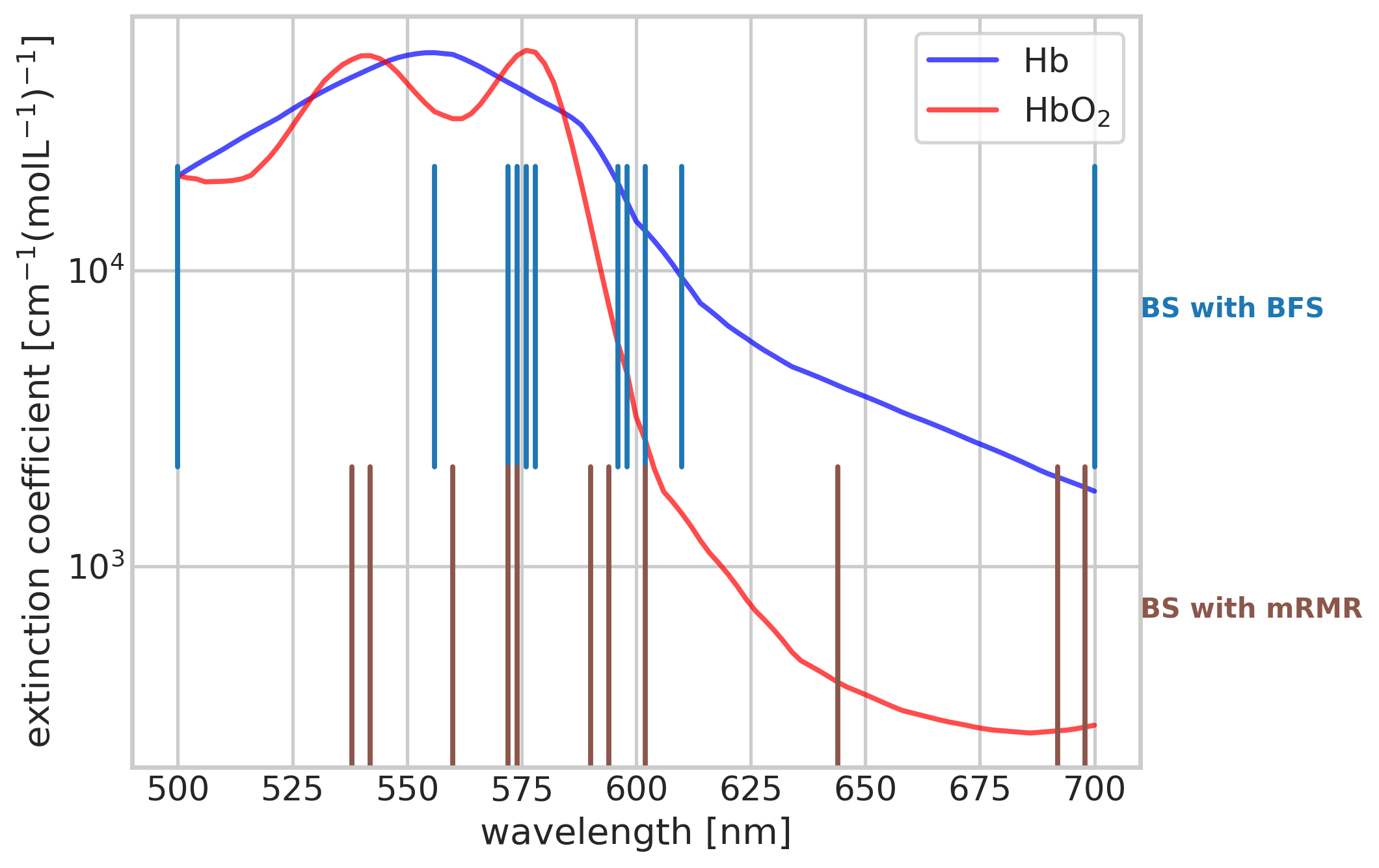}
}
\end{tabular}
\caption{(a) MAE of various band selection methods. Selected on the generic training set $X_g$ and evaluated on the test set $X_g^{test}$. \textit{Wrapper} methods (W) outperform \textit{filter} methods (F). The horizontal line represents the result when training and testing on all 101 bands. (b) The first 11 selected bands from the best \textit{wrapper} (BFS) and the best \textit{filter} method (mRMR) for the oxygenation estimation task. This particular number of bands was chosen as this is where the MAE achieved with BFS is lower than the baseline (101 bands).}
\label{fig:sfsmrmr_sao2_selections}
\end{figure}

Several popular \textit{filter} and \textit{wrapper} methods were evaluated. Publicly available code for all feature selection methods used in this manuscript is provided \href{https://github.com/FabianIsensee/FeatureSelection}{online}. Results for these methods on the oxygenation task are shown in Fig. \ref{fig:filterwrapper}. The selected bands for the best \textit{filter} and best \textit{wrapper} method are shown in Fig. \ref{fig:bsmethods}.

\begin{figure}[h!]
    \centering
    \includegraphics[width = 0.7 \columnwidth]{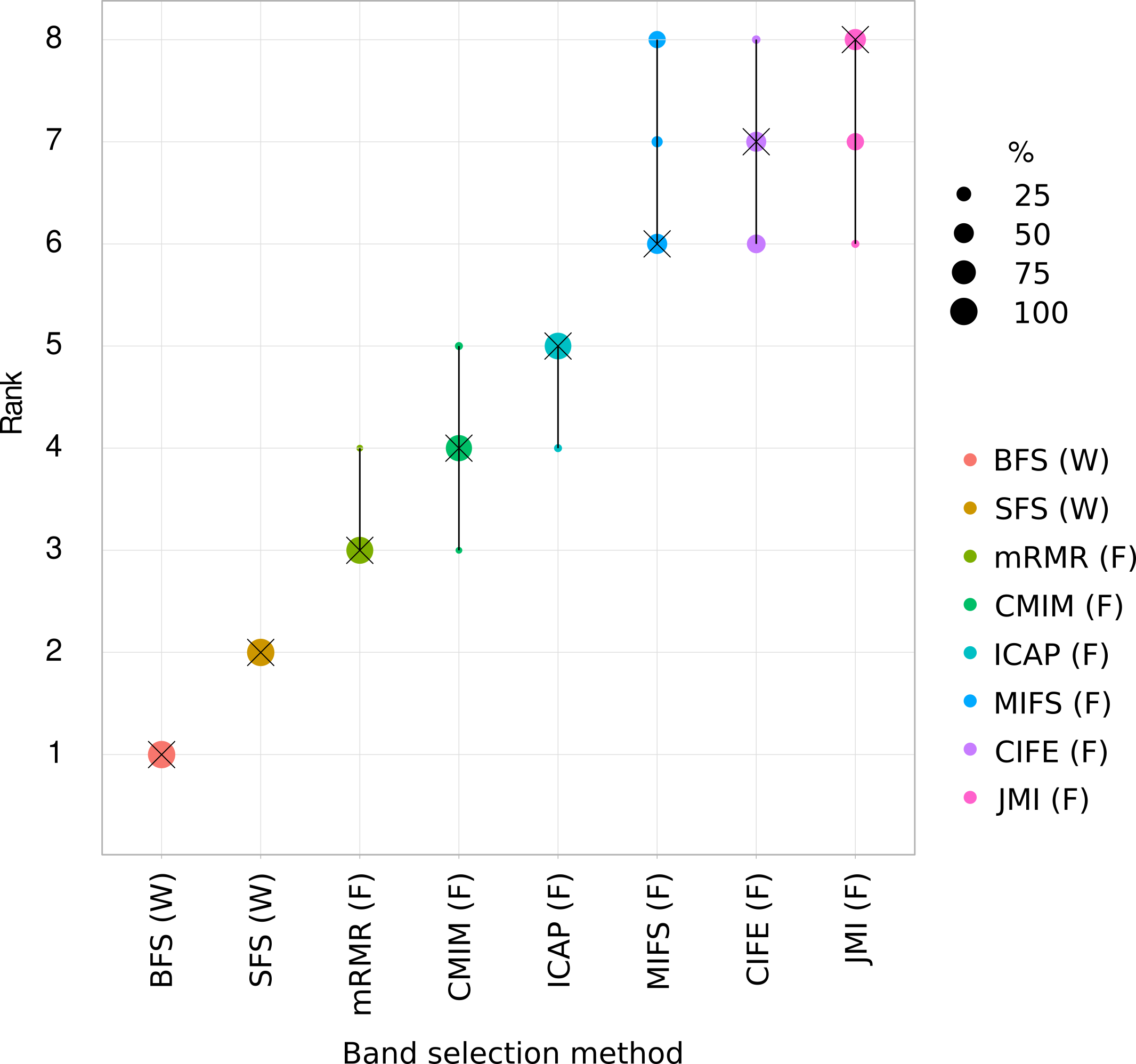}
    \caption{Ranking stability of different band selection methods across different number of bands (see Fig. \ref{fig:filterwrapper}). Here the rank of a method (1: best to 5: worst) is based on the MAE. Each method is color-coded, and the area of each blob at position ($A_i$, rank $j$) is proportional to the relative frequency ($A_i$) each method achieved rank $j$ for 1000 bootstrap samples. The median rank of each algorithm is indicated by a black cross and 95\% bootstrap confidence intervals across bootstrap samples are indicated by black lines.}
    \label{fig:ranking}
\end{figure}

A systematic robustness analysis, shown in Fig. \ref{fig:ranking} and performed with the \href{https://github.com/wiesenfa/challengeR}{\textit{ChallengeR}} tool developed by Wiesenfarth \textit{et. al.} \cite{wiesenfarth2021}, shows that the best method across different numbers of bands is BFS.
Because \textit{wrapper} methods performed better than \textit{filter} methods and BFS ranks in first place for all different numbers of bands, it was selected as the standard band selection technique for the following experiments. 

\begin{figure}[h!]
\centering
\begin{tabular}{ccc}
\subfloat[MAE for different number of bands\label{fig:redundant_mse}]{ %
\includegraphics[width=0.420\columnwidth]{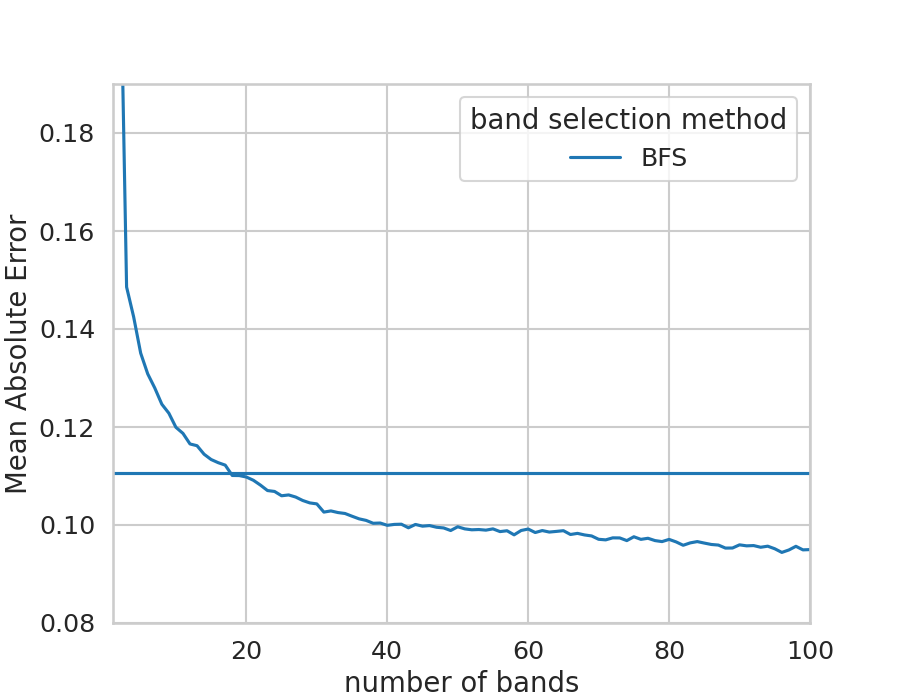}
\label{fig:redundantplot}
}  &
\subfloat[Most selected bands represented as vertical lines\label{fig:redundant_mua}]{%
 \includegraphics[width=0.395\columnwidth]{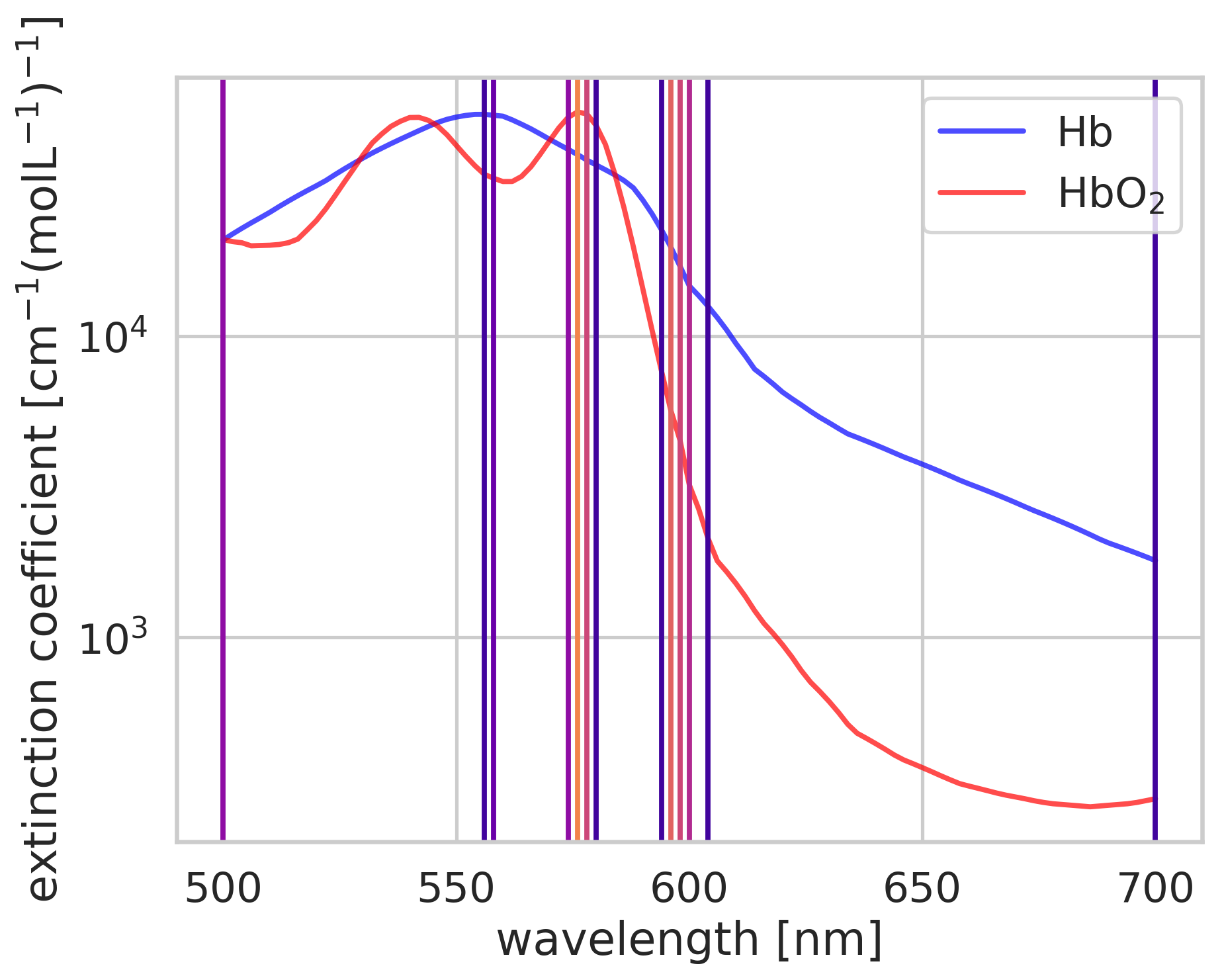}
 \label{fig:noisePlot}
}  &
\subfloat{%
  \includegraphics[width=0.094\columnwidth]{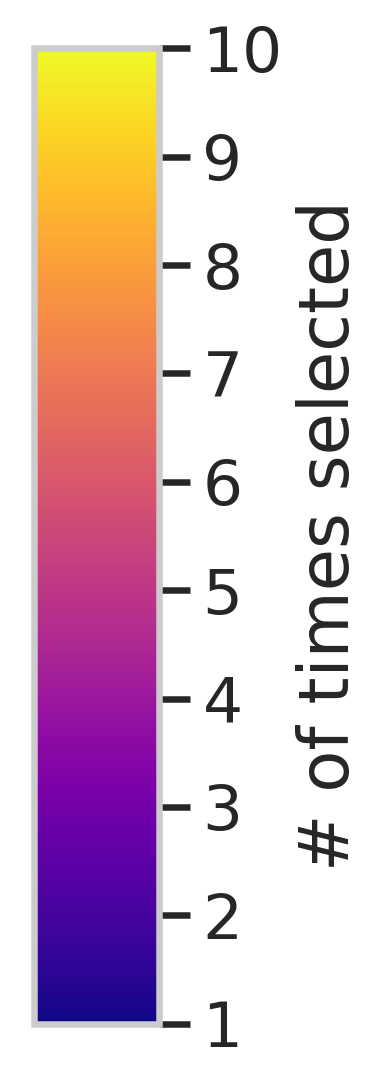}
  \label{fig:noiseStrips}
}
\end{tabular}
  \caption{Redundant band experiment. Each band was copied 10 times, each with its own noise applied to it, this allows BFS to select a band more than once. (a) MAE achieved with an increasing number of selected bands (BFS). The horizontal line indicates the MAE achieved when using all 101 bands. (b) The forty most relevant bands as selected by BFS in this experiment. The number of selections for each band is color coded. Hemoglobin extinction coefficients are plotted for reference.}
  \label{fig:redundantBands}
\end{figure}

\paragraph{Influence of redundant information on band selection results}
The band selection results from the last paragraph showed that BFS picked bands in close proximity (e.g. 578nm and 576nm). This suggests that there are few distinct places highly suited for oxygenation estimation and the algorithm is adding redundancy and thereby increasing its robustness to noise by selecting adjacent bands. To investigate this effect further we designed an experiment which gives the algorithm more freedom to build redundancy in the chosen bands. To this end, each band was duplicated ten times. Each duplicate was then augmented with additive Gausian noise (SNR 10). This corresponds to a maximum SNR of about 32 if all ten bands of a given wavelength were selected. Fig. \ref{fig:redundant_mse} shows the MAE achieved with an increasing number of selected bands; it can be seen that the baseline MAE (horizontal line) is reached with approximately 20 bands when selecting bands with BFS. 
Fig. \ref{fig:redundant_mua} shows the first 40 bands selected in this manner. The mean number of selections per band was $3.1$. The highest number of selections were performed for center wavelengths of 576nm (7), 596nm (6) 598nm (5) and 578nm (5), where the number of selections is given in parentheses.

\begin{figure}[h!]
\centering
\begin{tabular}{cc}

\subfloat[Influence of noise level on MAE]{%
 \includegraphics[width=0.5\columnwidth]{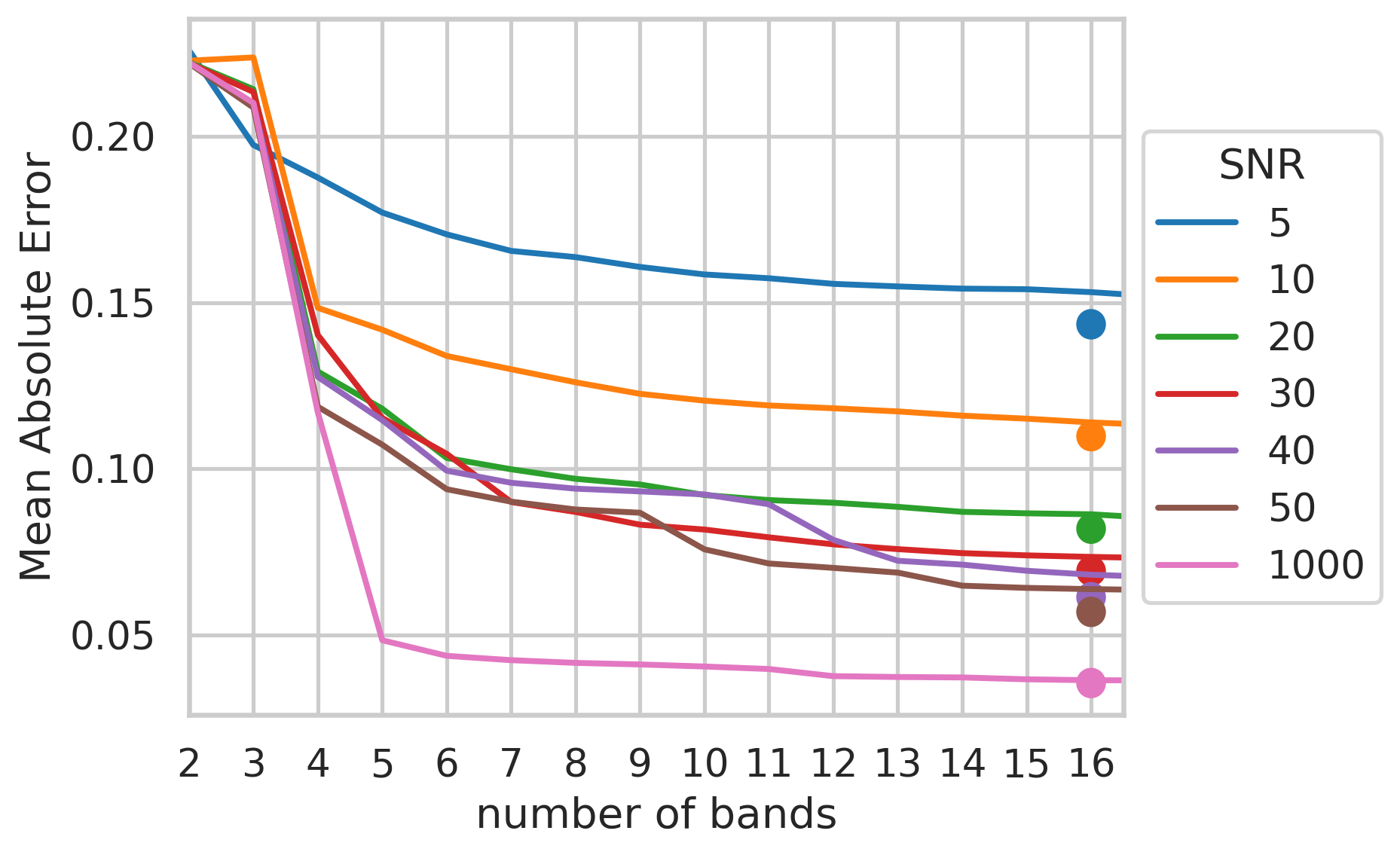}
 \label{fig:noisePlotBS}
}  &
\subfloat[Influence of noise level on band selections]{%
  \includegraphics[width=0.45\columnwidth]{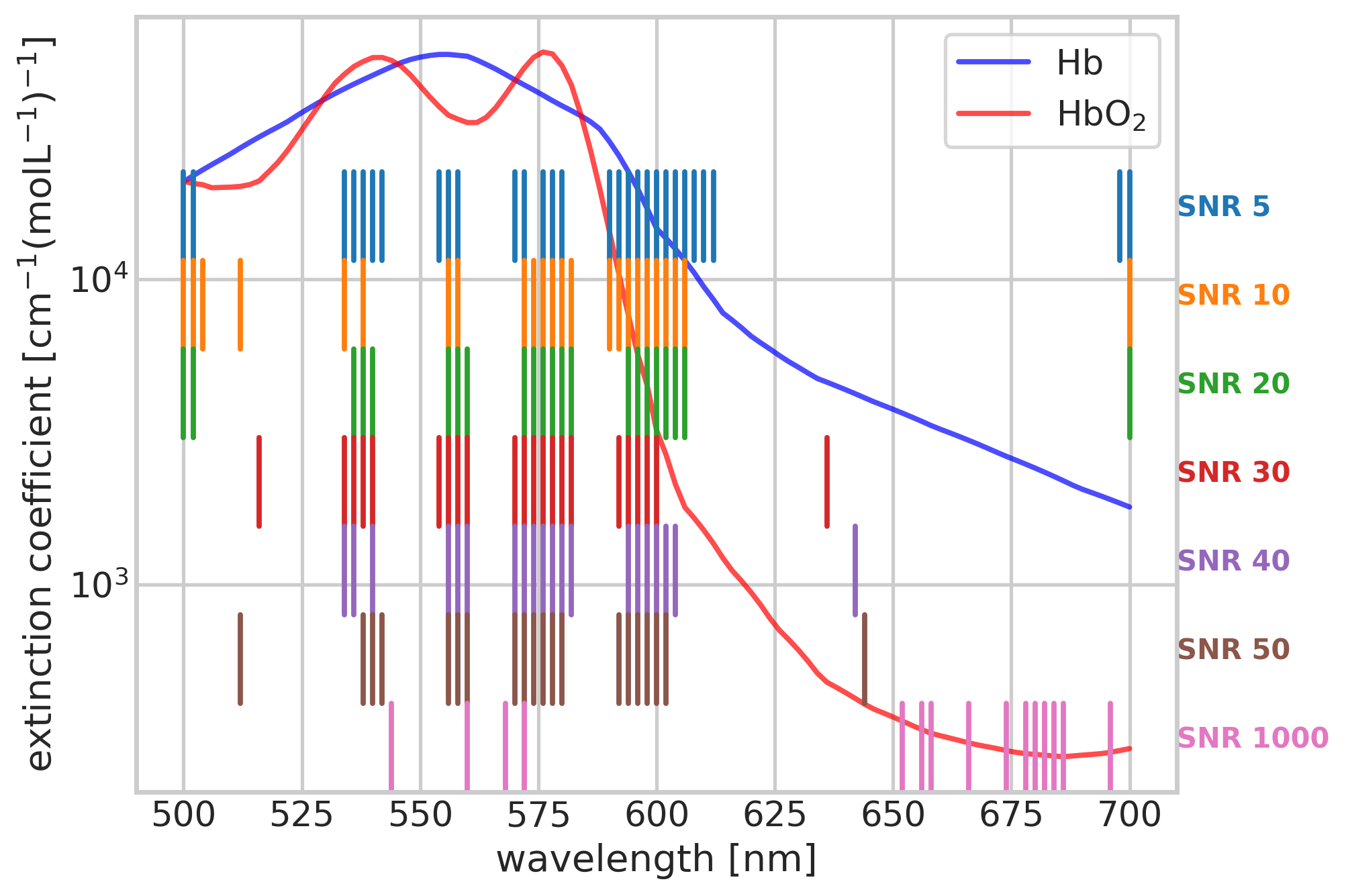}
  \label{fig:noiseStripsBS}
}
\end{tabular}
 \caption{(a) MAE on varying noise levels and number of selected bands, the circles show the minimum MAE achieved for each SNR. With 16 bands, performance for SNR=1000 is close to the best result indicated by the dot. (b) The stripes indicate the center wavelengths of bands for different noise levels. The number of bands shown here corresponds to the first combination to be within a margin of 0.01 compared to the \textit{best} achieved MAE (circles). Hemoglobin extinction coefficients are plotted for reference.}
\label{fig:noisePlots}
\end{figure}

\paragraph{Influence of noise on necessary number of bands}
This experiment aimed to investigate how many bands are needed and whether or not this number of bands is dependent on the expected SNR of the multispectral system. 
The SNR in the simulations varied between 5 (very high noise) and 1000 (virtually no noise). Note that unlike in the previous paragraph, bands are not duplicated and can be selected only once.
Fig. \ref{fig:noisePlotBS} shows how the SNR and number of selected bands influence the MAE. For noise levels of 5, 10, 20, 30, 40, 50, 1000 SNR, the minimum number of bands to be within a 0.001 margin of the best MAE for the SNR level was 29, 24, 22, 22, 20, 20 and 15 respectively. The \textit{best MAE} hereby refers to the minimum MAE achieved with the band set identified throughout this experiment, not the 101 bands baseline. Fig. \ref{fig:noiseStripsBS} shows the selected bands.

\begin{figure}[h!]
  \begin{center}
  \includegraphics[width=0.6\textwidth]{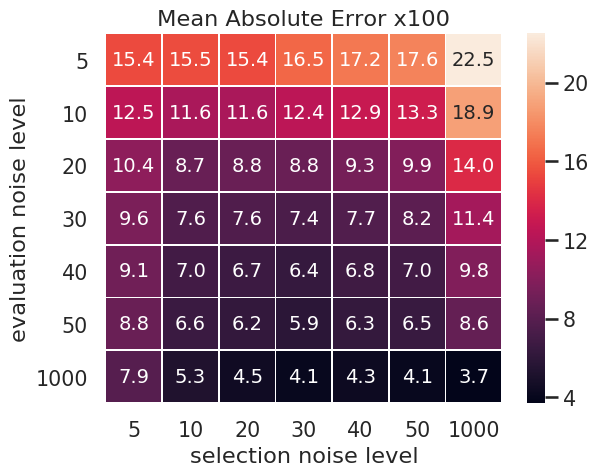}
  \end{center}
  \caption{Performance of our proposed method when using different noise levels for selecting and applying bands. The matrix shows how the MAE varies when selecting fifteen bands on one noise level and training and testing these selected bands on another noise level.}
  \label{fig:crossnoiseplot}
\end{figure}

Sources of noise, such as camera noise and model uncertainties, can be difficult to quantify or estimate in advance. We therefore investigated the effect of selecting bands with one noise level and evaluating on another one. We chose the best set of fifteen bands with SNR levels of 5, 10, 20, 30, 40, 50 and 1000 because at this threshold SNR=1000 was within 0.01 of its \textit{best} MAE. Evaluation was done using all combinations of bands and SNRs. Fig. \ref{fig:crossnoiseplot} shows how these factors interplay. As can be seen in the figure, bands selected at one noise level will not provide ideal results for another noise level. This is because high amounts of noise favor the addition of neighboring and thus redundant bands earlier on in the band selection process to make the regressor more robust, while low amounts of noise allow for earlier stratification into other wavelength regions.

\paragraph{Influence of domain shift on band selection}
The influence of the target domain on the band selection and the possibility to adapt to the target domain using the method presented in Sec. \ref{sec:da} are investigated in the following.
For this, a separate set of colon simulations $X_c$ with 15,000 simulated reflectances drawn from the tissue model specified in Table \ref{tab:colon} was created. 10,000 of these simulations were used to perform domain adaptation as described in Sec. \ref{sec:da} on $X_g^{\textrm{train}}$. The remaining 5,000 simulations were used as a test set. To form a data set adapted to colon tissue $X_{ac}$, 15,000 simulations were sampled with replacement using $X_g^{\textrm{train}}$ and the weights determined by KMM. 

\begin{table}[h!]
\centering
\caption{Parameter ranges for the colon tissue model and their usage in the simulation setup as described in \cite{wirkert_robust_2016}. Here $v_{\textrm{Hb}} \lbrack \% \rbrack$ represents the blood volume fraction, $s$ the blood oxygen saturation, $a_{mie}$ the reduced scattering coefficient at $500 nm$, $b_{mie}$ the scattering power, $g$ the tissue anisotropy, $n$ the refractive index and $d$ the tissue thickness.}
\label{tab:colon}

\begin{tabular}{p{1.1cm}c c c c c c c}
\hline
      & $v_{\textrm{Hb}}\lbrack \% \rbrack$           & $s \lbrack \% \rbrack$  & $a_{\text{mie}} \lbrack \si{cm^{-1}}\, \rbrack$ & $b_\text{mie}\unit{[a.u.]}$          & $g\unit{[a.u.]}$      & $n\unit{[a.u.]}$         & $d \lbrack\si{cm}\rbrack$    \\ 
\hline \hline
layer 1
: & $0-10$ & $0-100$ & $8.7-29.1$ & 1.289 & $0.80-0.95$ & 1.36 & $0.06-0.11$  \\ 
layer 2
: & $0-10$ & $0-100$ & $8.7-29.1$ & 1.289 & $0.80-0.95$ & 1.36 & $0.04-0.08$  \\ 
layer 3
: & $0-10$ & $0-100$ & $8.7-29.1$ & 1.289 & $0.80-0.95$ & 1.38 & $0.04-0.06$  \\ \hline \hline
\multicolumn{8}{l}{$\mu_a(v_{\text{Hb}}, s, \lambda) = v_{\text{Hb}} (s\cdot \epsilon_{\text{HbO2}}(\lambda) + (1-s)\cdot\epsilon_{\text{Hb}}(\lambda))\cdot \ln (10)\cdot150\si{g.L^{-1}}\cdot(\num{6.45e4}\si{g.mol^{-1}})^{-1}$} \\
\multicolumn{8}{l}{$\mu_s(a_{\text{mie}}, b, \lambda) = \frac{a_{\text{mie}}}{1-g}(\frac{\lambda}{500\si{nm}})^{-b_{\text{mie}}}$}  \\ \hline                 
\multicolumn{8}{l}{simulation framework: GPU-MCML\cite{alerstam_next-generation_2010}, $10^6$ photons per simulation}                                                \\                  
\multicolumn{8}{l}{simulated samples: $\num{2e4}$}                                                \\
\multicolumn{8}{l}{sample wavelength range $[\lambda_{min}-\lambda_{max}]$: 450\si{nm}-720\si{nm}, stepsize 2\si{nm}}\\
\hline
\end{tabular}
\end{table}

\begin{figure}[h!]
\centering
\begin{tabular}{cc}

\subfloat[MAE of different data sets\label{fig:da_mae}]{%
 \includegraphics[width=0.45\columnwidth]{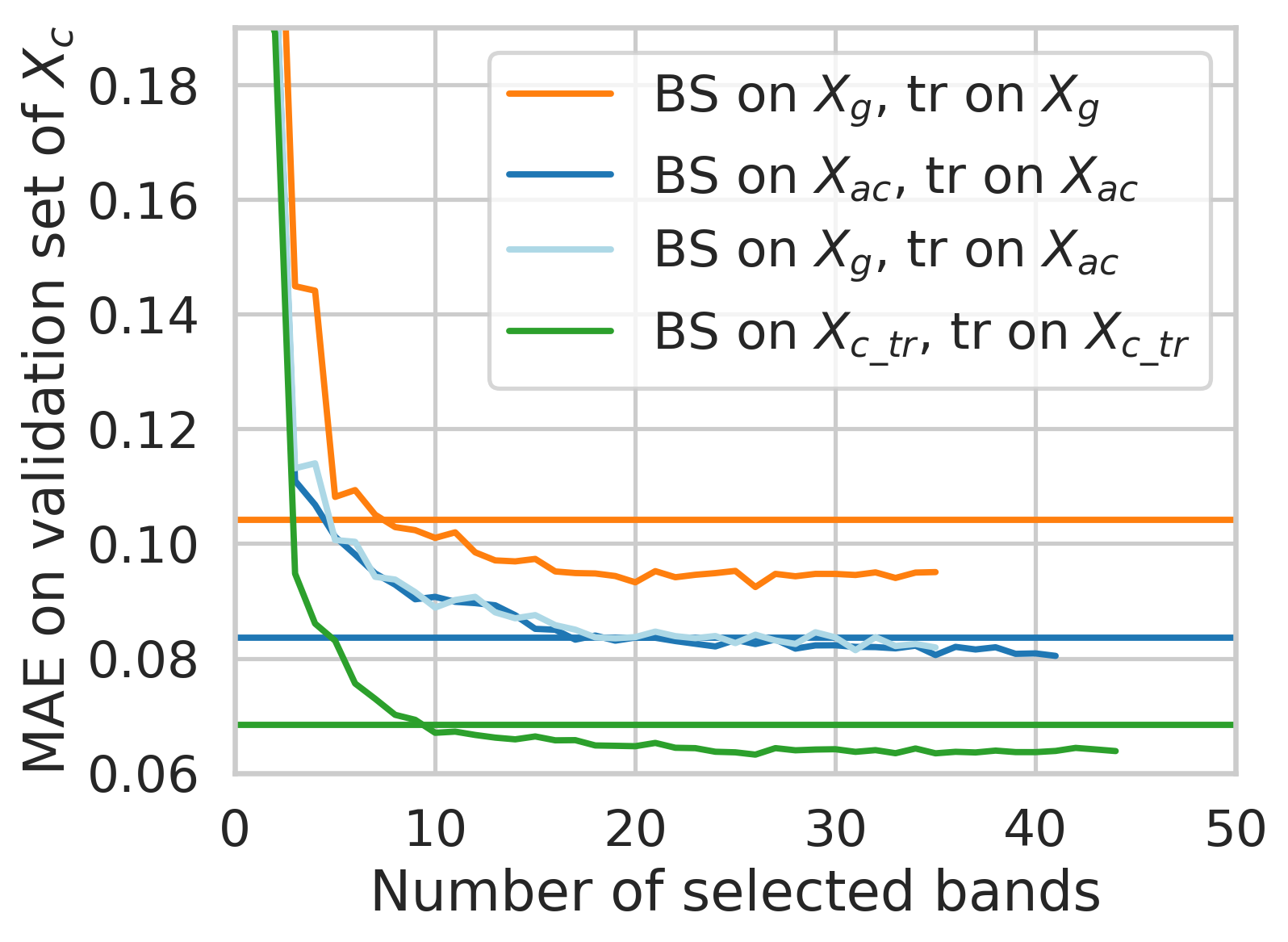}
}  &
\subfloat[Selected bands for different source domains\label{fig:da_selection}]{%
  \includegraphics[width=0.5\columnwidth]{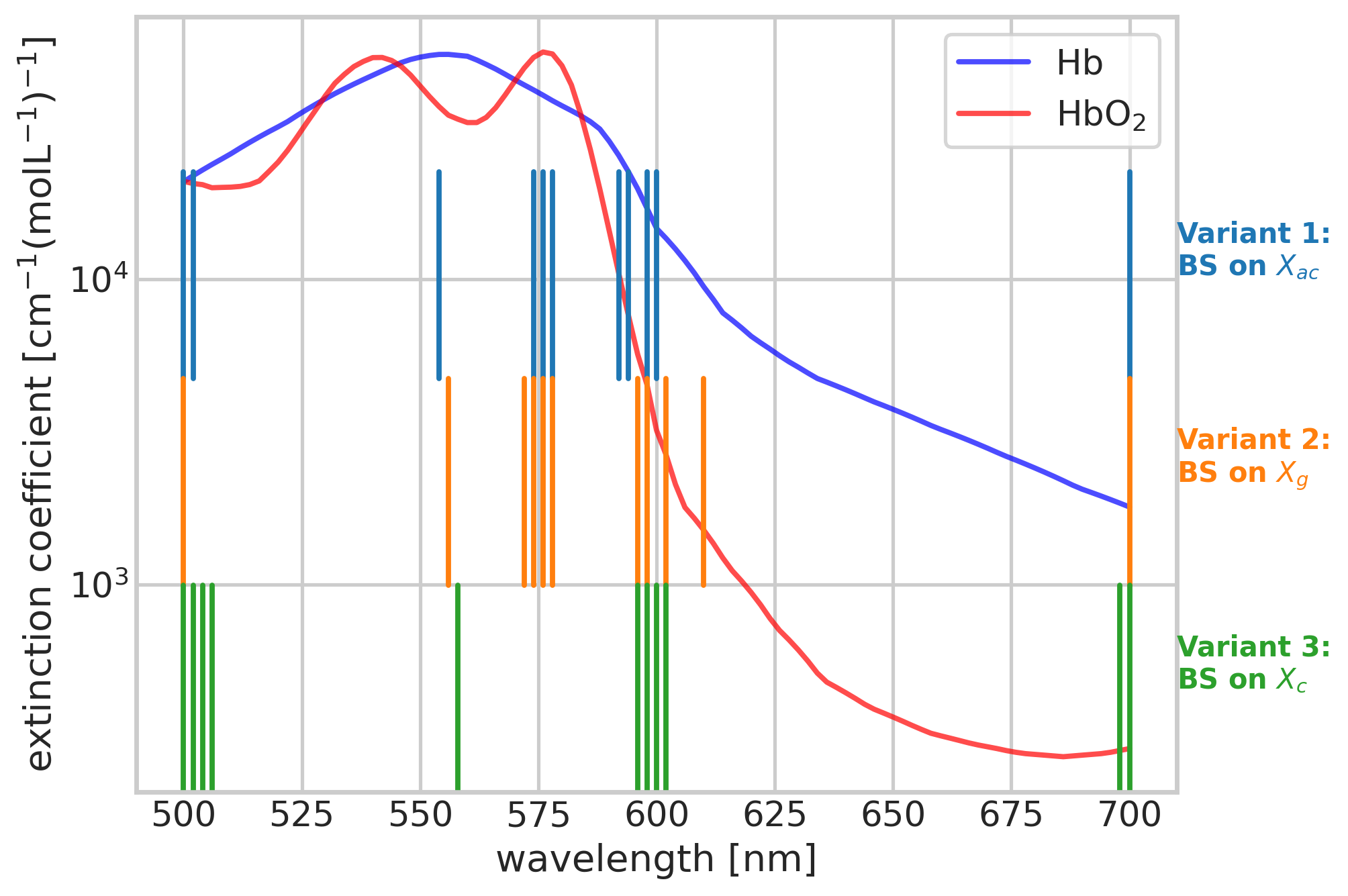}
}
\end{tabular}
 \caption{(a) Comparison of bands selection (BS) using domain-specific $X_{ac}$ (dark blue) and unadapted simulations $X_g$ (orange). As an upper bound, the result of selecting on the simulated colon target domain $X_c$ (green) is shown. The light blue line shows the result of selecting bands based on $X_g$ but training a regressor with data from $X_{ac}$. All experiments were evaluated on the test set of the target domain ($X_c$). The horizontal lines are the results for training and testing on all 101 bands of the respective training set (no band selection). (b) The ten first bands selected for each of the different source domains. Hemoglobin extinction coefficients are plotted for reference.}
 \label{fig:da_simcolon}
\end{figure}

Fig. \ref{fig:da_mae} shows how selection from domain-specific simulations compares to selection from unadapted data using BFS for band selection. For each experiment we indicate what data the bands were selected on (\textit{BS on}) and what data was used for training the regressor (\textit{tr on}). Evaluation was always done on the test set of the target domain $X_c$. We show results using general data $X_g$, domain adapted data $X_{ac}$ and target domain data $X_c$ (train) as source domain. Herewith, the results for $X_g$ should be interpreted as a lower bound and the results for $X_c$ are an upper bound. Note that in real-world scenarios annotated data from the target domain is usually not available, rendering band selection on $X_c$ impossible. As can be seen in Fig. \ref{fig:da_mae} (blue line), domain adaptation can indeed close the domain gap to some extent. While band selection did improve the MAE to below the respective baseline for $X_g$ and $X_c$, a considerable improvement was not observed for $X_{ac}$. Across all numbers of selected bands, domain adaptation yielded better results than just using $X_g$. The light blue line shows the MAE when selecting on $X_g$ but training on $X_{ac}$. This result indicates that, at least for the present data set, features may not need to be re-selected after domain adaptation. This is supported by the selected bands on each source domain depicted in Fig. \ref{fig:da_selection}.

\subsection{\textit{In vivo} validation}
\label{sec:invivo}
The \textit{in vivo} validation investigates how well bands selected from the (domain-specific) simulations can estimate oxygenation on real data. Images from \cite{lu_framework_2015} were taken for this evaluation. They encompass 8 hyperspectral images of head and neck tumors in a mouse model, captured by a Maestro
(PerkinElmer Inc., Waltham, Massachusetts) imaging system. This system records hyperspectral images from 450-900nm with a FWHM of 20nm. Green fluorescence protein (GFP) was used to identify tumorous tissue. For more details on the imaging process and the reference generation please refer to \cite{lu_framework_2015}. From the 8 mouse tumor images, five were used for algorithm fine-tuning and three were reserved for the final validation of the band selection results. For the following analysis, bands between 500-700nm were considered, as only in these ranges a close match between simulations and measurements could be established.

\begin{figure}[h!]
\centering
\includegraphics[width=\textwidth]{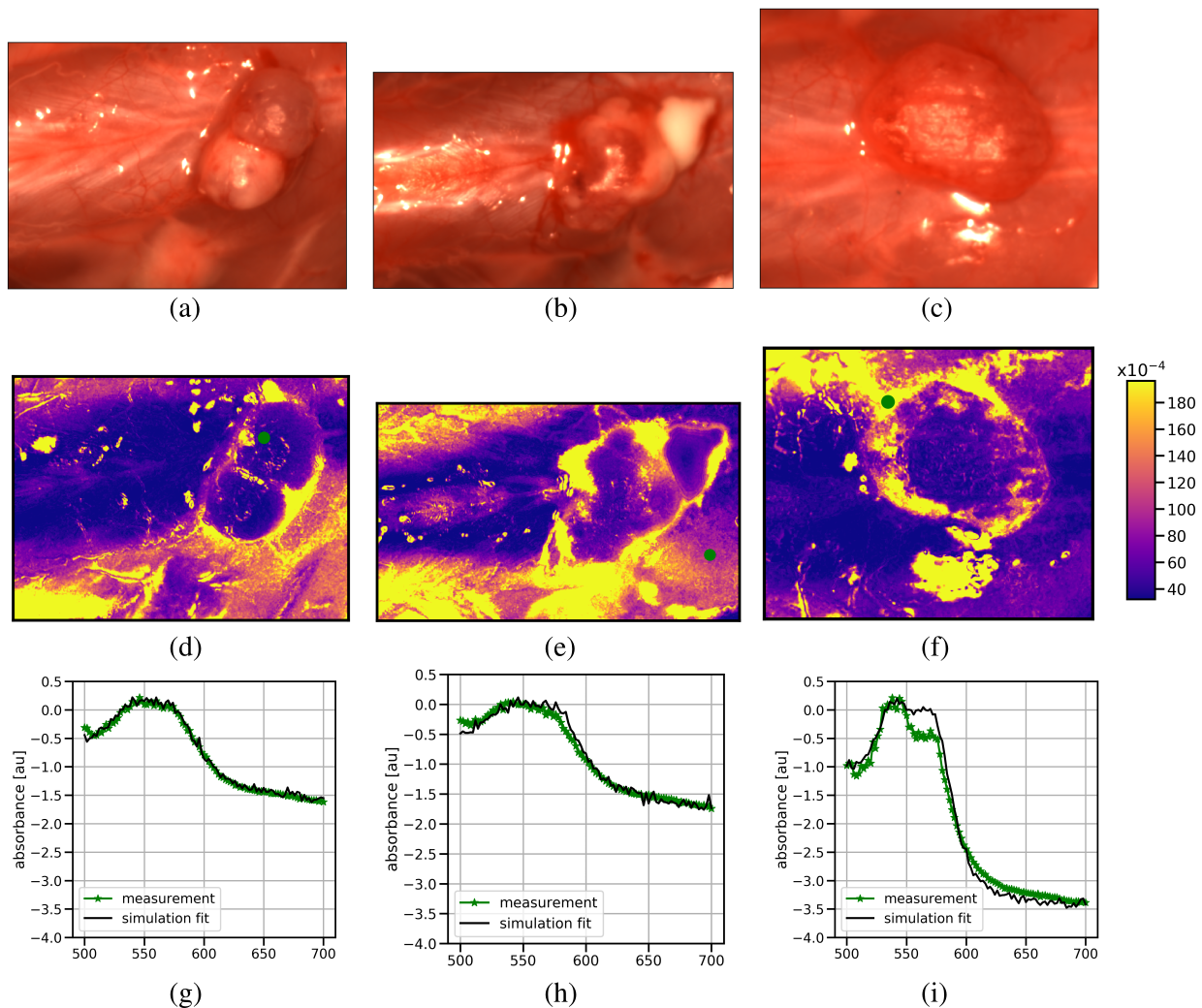}
 \caption{Reconstructed RGB (a,b,c) and fit error (d,e,f) for three training images. The fit error for each pixel is the smallest MSE compared to simulated data from $X_g$. (g,h,i) show the best simulation fit for the green points in the images above. They were selected to show a good (g), bad (h) and very bad (i) fit. }
\label{fig:image_fit}
\end{figure}

\paragraph{Matching simulations and real measurements}
We investigated how the simulations fit to the real mouse measurements for all pixels in the training images. For this purpose, the real measurements $I_{(i, j)}$ at spatial location $(i,j)$ and simulated reflectances $X_g$ were normalized as described in Sec. \ref{subsec:data-normalization}. This yielded the normalized real measurements {$\bar{I}_{(i, j)}$} and the set of normalized simulated spectra $\bar{X}_g$. Furthermore, the nearest neighbor simulation {$\boldsymbol{\bar{r}_{nn}}$} to each normalized measurement {$\bar{I}_{(i, j)}$} was determined by choosing the simulated reflectance {$\mathbf{\bar{r}}_i \in \bar{X}_g$} with the minimum mean squared error (MSE) to {$\bar{I}_{(i, j)}$}, that is:
$$(\boldsymbol{\bar{r}_{nn}})_{(i,j)} = \argmin_{\boldsymbol{\bar{r}}} \left( \frac{1}{N_s} \sum_{k=1}^{N_s} ((\bar{I}_k)_{(i, j)} - \bar{r}_k )^2\right); \hspace{0.5cm} \mathbf{\bar{r}}\in \bar{X}_g$$
We refer to the MSE between $\boldsymbol{\bar{r}_{nn}}$ and measurements $\bar{I}_{(i, j)}$ as \textit{fit error}. The median fit error was $6.6\times10^{-5}$. To provide the reader with an estimate of the distance between measurements and simulations, we plot exemplary images in Fig. \ref{fig:image_fit}.
To understand how the fit error varies within an image, we report three of the training images as reconstructed RGB color images (reconstructed from hyperspectral measurements) and as fit errors in Fig. \ref{fig:image_fit}. It can be seen that the regions of the images with the biggest fit error correspond to regions around specular reflections, or dark regions visible on the reconstructed RGB images.

\paragraph{Qualitative comparison of band selection methods}
The \textit{in silico} methods evaluated the band selection in a quantitative, although simulated environment. In this experiment we compared selected methods in an \textit{in vivo} setting. The guiding assumption was that the bands resulting from the band selection algorithms should be able to reproduce (and maybe improve upon) the full 101 band oxygenation estimate.

\begin{figure}[h!]
    \centering
     \includegraphics[width=0.65\textwidth]{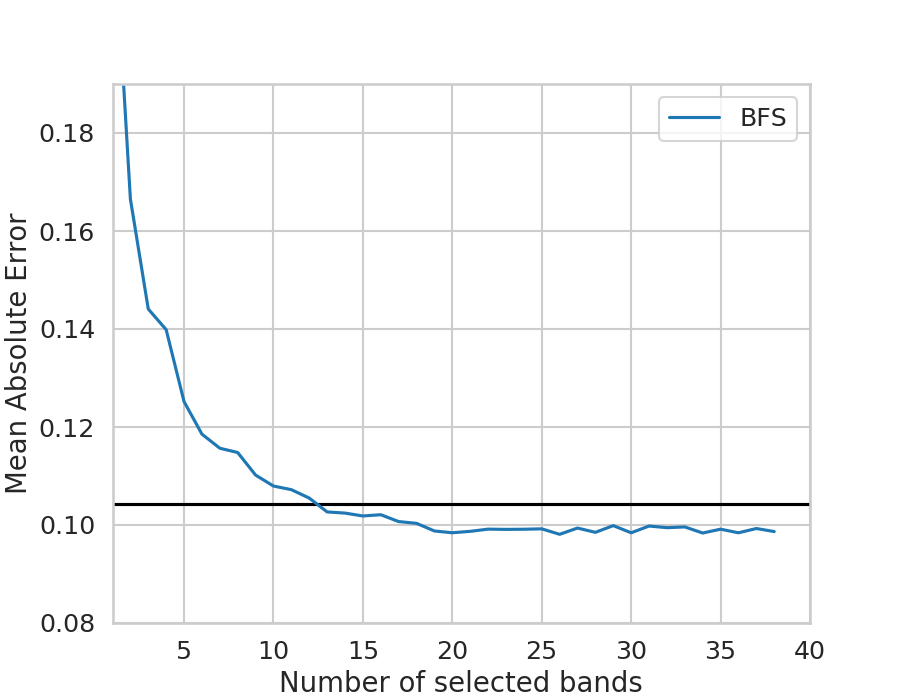}
     \caption{Band selection results on domain-specific data $X_{am}^{train}$ for the \textit{in vivo} experiment. Band selection was performed using BFS on the training set $X_{am}$ and the MAE are reported for the test set $X_{am}^{test}$. The horizontal line indicates the MAE achieved with all 101 bands. The best result was obtained with 20 bands. These bands are then used for the \textit{in vivo} data. Note that the results displayed in this figure are based on domain-adapted simulations only and do not provide a performance estimate for the \textit{in vivo} data.}
     \label{fig:xam_selection}
\end{figure}

We focused on BFS, the best performing \textit{wrapper} method. Experiments in this section are based on 15,000 domain adapted measurements as a training set $X_{am}^{train}$ and another 15,000 domain adapted measurements as a test set $X_{am}^{test}$, both sampled from $X_g$ using weights from our domain adaptation approach. We selected bands on the training set, then trained a regressor on the training data with the selected bands and finally evaluate the regressor on the test set. The results for the band selection on domain-adapted \textit{in silico} data are shown in Fig. \ref{fig:xam_selection}. As can be seen in the figure, the baseline MAE can be reproduced with 13 bands while the best MAE is achieved with 20 bands. Note that in this setup the band selection algorithm is agnostic to the real measurements as it is presented only domain-specific generic simulations. This approach allows us to select bands that are specific to some domain without having ground truth oxygenation values for the measurements. 

\begin{figure}[h!]
\centering
\includegraphics[width=\textwidth]{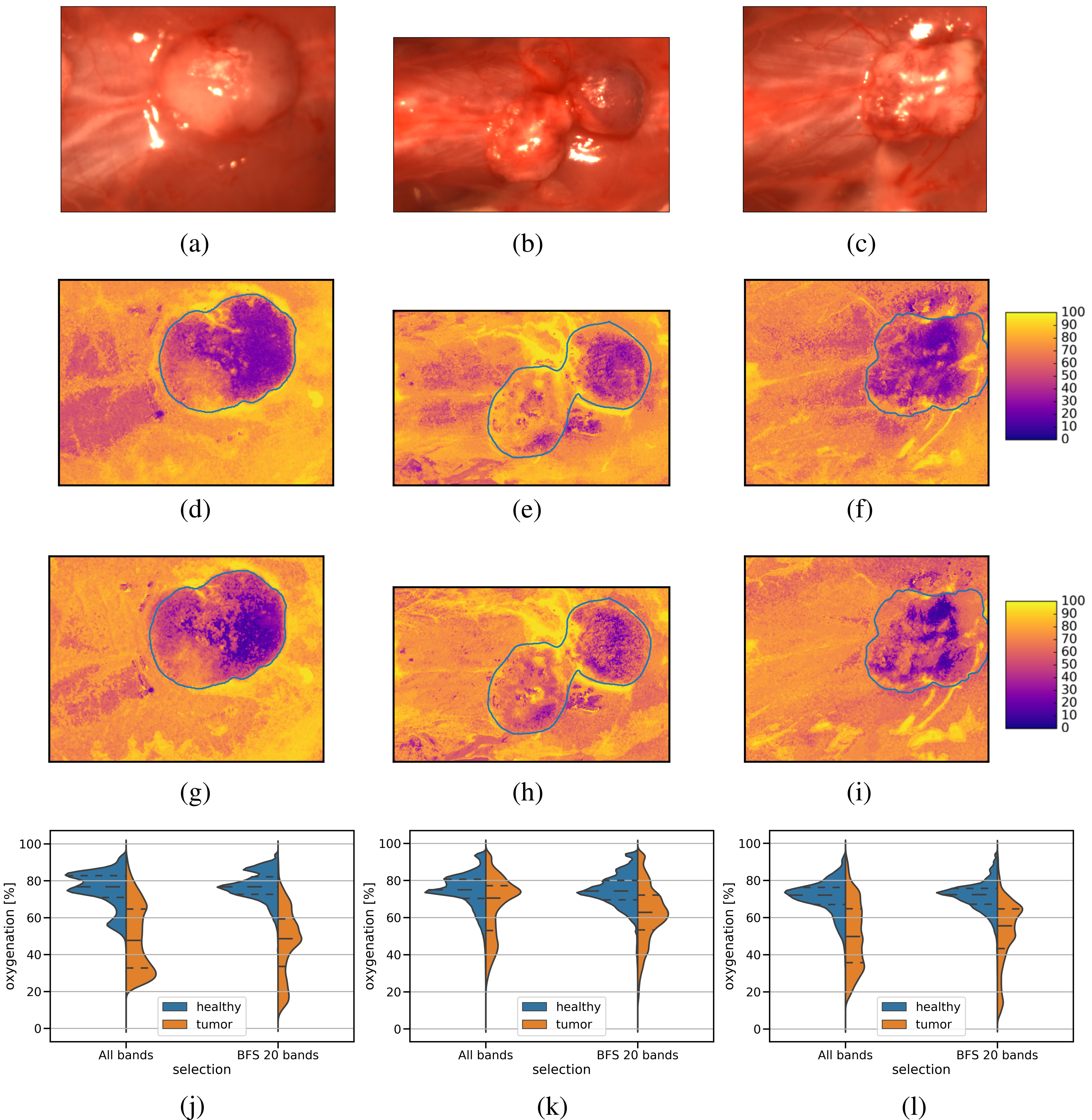}
 \caption{Reconstructed RGB (a,b,c) of the test images. (d), (e), (f) show the oxygenation estimate using all 101 bands. (g), (h), (i) show the oxygenation estimate using only 20 bands selected with BFS. (j), (k), (l) show the violin plots, both for tumorous and benign regions. Tumorous regions were identified via GFP and are indicated by the blue outline in the images. The color bars on the right show the percentage of oxygenation.}
\label{fig:invivo_test}
\end{figure}

\paragraph{Final evaluation on test images}
In this final evaluation, we compared the BFS result using the 20 selected bands from the previous section with the 101 band oxygenation result. For this evaluation, three previously unused tumor images were used. See Fig. \ref{fig:invivo_test} for a side by side comparison on the three images and violin plots of the tumor and non-tumor oxygenation results.

\section{Discussion}
\label{sec:dscussion}
The contributions of this paper can be summarized as follows. Firstly, we are - to our knowledge - the first to investigate whether a small subset of bands, selected using highly accurate simulations, can yield high-quality results on the clinically relevant task of oxygenation estimation. Secondly, we proposed the first method to select bands depending on the domain under investigation (here: head and neck tumor), without the need for (often unavailable) labeled \textit{in vivo} data. The approach relies on selecting bands from a large generic database of labeled \textit{in silico} reflectances generated with MC methods. Previously proposed domain adaptation techniques were employed to select simulations which mimic the tissue of interest. Finally, we compared a large number of state-of-the-art band selection techniques as part of our framework. In the following, the results are discussed in light of the \textit{in silico} and \textit{in vivo} experiments.

\paragraph{\textit{In silico} experiments}
The \textit{in silico} experiments investigated several aspects of the band selection algorithm in a controlled environment. The key findings were that (a) \textit{wrapper} outperform \textit{filter} methods, (b) 10 bands give a performance comparable to 101 bands in this setting, (c) selecting more bands can yield better than baseline MAE in the case of \textit{wrapper} methods, (d) the algorithm adds redundant information by choosing neighboring bands, (e) different noise levels impact band selection results with higher noise levels favoring redundancy early in the selection process, and (f) band selection is beneficial even when adding additional complexity to the experimental pipeline by incorporating the proposed domain adaptation technique.

\textit{Wrapper} methods consistently outperformed \textit{filter} methods in our experiments. This is probably explained by being more ``end-to-end'' than the \textit{filter} methods, which are agnostic to the downstream regression pipeline (here a random forest regressor). When looking at the selected bands, it can be noticed that the \textit{filter} methods select bands more broadly across the spectrum of possible wavelengths. In contrast, \textit{wrappers} first pick bands in places with high differences between oxygenated and deoxygenated hemoglobin while later adding redundancy via neighboring bands. Adding redundant bands can be beneficial in the noise model used here (see below). This benefit could however not be captured by the heuristics employed in the \textit{filter} methods. \textit{Filter} methods typically penalize high mutual information between features in the selected set, thus refusing to select neighboring bands. 

Our choice of normalization method is a design choice that needs to be discussed. In this work the band selection algorithms could influence the normalization because the nature of the normalization method used here required a re-normalization of the data whenever a feature was removed or added. To investigate the influence of the normalization method, we changed it to an image quotient norm \cite{styles_quantitative_2006} (data not shown). With the image quotient norm, this re-normalization step is unnecessary and therefore eliminates the possibility for the band selection algorithms to jointly optimize the regressor and the normalization. In our \textit{in silico} experiments, the image quotient norm resulted in higher overall MAEs, the algorithms never surpassed the 101 baseline and a less pronounced effect of the domain adaptation was observed. Furthermore, we observed that the relative performance of the different methods is independent of the choice of normalization. Interestingly, in the \textit{in vivo} experiments, image quotient normalization led to almost exact reproduction of the 101 band baseline result with 16 bands. A potential loss in MAE as observed in \textit{in silico} experiments is likely, but could not be determined \textit{in vivo} due to the lack of ground truth oxygenation. We leave additional evaluation of the influence of normalization strategies to future work.

We further explored the benefits of redundancy by allowing the \textit{wrapper} algorithms to select bands multiple times. Hereby, we could observe that only few very specific places are selected (see Fig. \ref{fig:redundantplot}). After a certain amount of bands the algorithm seems to focus on redundancy. 

This observation is in line with observations made by Guyon et al. \cite{guyon2003introduction} who report that redundant but noisy features are beneficial if used jointly to average out the noise. This behavior is certainly linked to the independent multiplicative noise model applied to the simulated data and might change if the noise model accounted for inter-band dependencies. In our experiments using different SNR settings, it could be observed (Fig. \ref{fig:noiseStrips}) that, although the number of bands needed increases slightly when noise is added, the same general spectral regions are selected (except for cases when very low noise levels are applied). Due to the varying noise levels, the \textit{wrapper} method will sometimes start adding redundant bands sooner, sometimes later. In this context, low noise levels favor more diverse regions within the studied spectral range because the robustness to noise via redundancy is not needed, and the regressor can therefore fit the training data better. This results in suboptimal feature sets when transferring them between different noise levels.

As expected, the chosen bands are mostly located close to areas of high difference in oxygenated and deoxygenated hemoglobin absorption.
The bands selected by the algorithm are thus predominantly within a small spectral range of 530-610nm. A reason for this might be that spectra within this range are disturbed less by variation in other confounding factors such as scattering, which can be assumed constant within small ranges.

The domain shift experiment demonstrated the effectiveness of the domain adaptation technique used in this paper. Compared to training on the general simulation data, the domain gap could partially be closed (see horizontal lines in Fig. \ref{fig:da_simcolon}). Using band selection in the context of a domain shift did not negatively impact results. Across all experiments done in this section, around 20 bands yielded the best oxygenation estimates. Interestingly, the selected bands are quite similar, especially between selection on $X_g$ and $X_{ac}$. This is underlined by the light blue line in Fig. \ref{fig:da_simcolon} which shows that features selected on $X_g$ are equally well suited as those selected on domain-adapted data $X_{ac}$, indicating that feature sets selected on $X_g$ may be broadly applicable across different domains.

The spectral range of the simulated camera was restricted to 500-700nm and the FWHM was set to 20nm. In future work, we will investigate how these results generalize to different camera setups and normalization schemes. \textit{Embedded} feature selection methods such as the one proposed in \cite{cong_deep_2015} could also be explored. Specifically, Cong et al. select features by adding a $\ell_1$ regularization term to the loss function of a support vector machine (SVM). This naturally leads to a feature selection by driving weights of less important features down to 0. However, \textit{embedded} approaches are considerably more restrictive compared to the \textit{wrapper} methods employed here because they are limited to machine learning regressors that can incorporate such an additional regularization term.

\paragraph{\textit{In vivo} experiments}
The \textit{in vivo} study relied on images of 8 head and neck tumors in a mouse model. We investigated band selection from 101 bands ranging from 500-700nm within the context of oxygenation estimation. The key findings were that (a) domain-adapted simulations and measurements show good alignment, (b) experiments on data adapted to the target domain show that BFS closely reproduces the MAE of the 101 band results for just 13 selected bands and achieved the lowest MAE with 20 bands, and (c) using these 20 selected bands for a qualitative analysis of the test images reveals similar predictions to the 101 bands estimate. 

Interestingly, the contrast within the tumor regions seems to be sharper for the 20 bands selected by BFS. However, due to the lack of ground truth oxygenation data, quantitative evaluation of the band selection result is not possible for this experiment. Therefore, we are unable to definitively determine whether the MAE of the selected 20 bands is indeed lower than the 101 bands estimate, as suggested by our \textit{in silico} experiments with domain adapted data (see Fig. \ref{fig:xam_selection}).

\paragraph{Conclusion}
We presented a method enabling task-specific band selection based on highly accurate simulations. \textit{In vivo} and \textit{in silico} results suggest the band selection can be performed purely \textit{in silico}, which greatly increases flexibility and reduces costs for selecting appropriate bands.

\paragraph{Data availability:} Data underlying the results presented in this paper are not publicly available at this time but may be obtained from the authors upon reasonable request.

\paragraph{Ethics:} The appropriate ethics approval for the animal experiments was obtained by the authors of the publication that describe the dataset \cite{lu_framework_2015}.

\paragraph{Acknowledgments:} The authors would like to acknowledge support from the European Union through the ERC starting grant {\footnotesize COMBIOSCOPY} under the New Horizon Framework Programme under grant agreement ERC-2015-StG-37960. The research is further supported in part by NIH grants CA156775, CA204254, and HL140135. Part of this work was funded by Helmholtz Imaging, a platform of the Helmholtz Incubator on Information and Data Science.

\bibliography{main}

\end{document}